\documentclass[9pt,twocolumn,twoside]{osajnl_copy}

\journal{ol} 

\setboolean{shortarticle}{true}

\title{Frequency ratio of an $^{115}$In$^+$ ion clock and a $^{87}$Sr optical lattice clock}

\author[*]{Nozomi Ohtsubo}
\author{Ying Li}
\author{Nils Nemitz}
\author{Hidekazu Hachisu}
\author{Kensuke Matsubara}
\author{Tetsuya Ido}
\author{Kazuhiro Hayasaka}

\affil{National Institute of Information and Communications Technology, Nukuikitamachi, Koganei, Tokyo 184-8795, Japan}

\affil[*]{Corresponding author: email: ohtsubo@nict.go.jp}

\doi{\url{https://doi.org/10.1364/OL.404940}}

\begin{abstract}
We report on the first frequency ratio measurement of an $^{115}$In$^{+}$ single ion clock and a $^{87}$Sr optical lattice clock.
A hydrogen maser serves as a flywheel oscillator to measure the ratio by independent optical combs.
From 89~000 seconds of measurement time, the frequency ratio $f_{\rm{In}} / f_{\rm{Sr}}$ is determined to be 2.952~748~749~874~863~3(23) with 7.7$ \times $10$^{-16}$ relative uncertainty.
The measurement creates a new connection in the network of frequency ratios of optical clocks. 
\end{abstract}

\begin{document}

\maketitle

Recent progress of optical clocks based on trapped ions and neutral atoms has made their uncertainties much smaller than those of the best cesium frequency standards \cite{Hong_2016,Riehle2018,Lodewyck2019}.
Now a new definition of the SI unit of time using the optical clocks has become realistic. 
The Consultative Committee for Time and Frequency (CCTF) drafted a roadmap to the redefinition of the SI second, where as a prerequisite to the new definition it requires measurements of frequency ratios \cite{Rosenband2008,Godun2014,Matsubara2012,Tyumenev2016,Nemitz2016,Ohmae2020,BACON2020} between different atomic transitions as well as the reproducibility of the clocks in different laboratories, both with uncertainty matching that of the best clocks \cite{Riehle2018}.

The indium ion ($^{115}$In$^+$) is one of the original candidates in the first proposal for single-ion clocks with systematic uncertainties at the 1$0^{-18}$ level \cite{Dehmelt1983}.
In$^+$ has an alkaline-earth-like electron configuration (as does Al$^+$), with a $^1$S$_0$--$^3$P$_0$ clock transition frequency that has a low sensitivity to magnetic fields.
It has the highest clock frequency of all presently investigated optical clocks \cite{Riehle2018}.
Such high transition energies give it a very small sensitivity to blackbody radiation (BBR), which causes problematic frequency shifts in most other clocks, including optical lattice clocks.
Direct state detection and laser cooling on the $^1$S$_0$--$^3$P$_1$ transition are a further advantage for an $^{115}$In$^+$ clock.
In lighter elements like $^{27}$\!Al$^+$, the weaker hyperfine mixing limits the scattering rate to little more than a kHz.
Although an Al$^+$ clock now reports a systematic uncertainty below 10$^{-18}$ \cite{Brewer2019}, the quantum state needs to be transferred to another ion by quantum logic spectroscopy.
The $^{115}$In$^+$ quantum state can be read out by a straight-forward shelving method due to a $^3$P$_1$ lifetime of only 440 ns.
The electric quadrupole moments of the two clock states are orders of magnitude smaller than those for the $^2$S$_{1/2}$--$^2$D$_{5/2}$ quadrupole transitions of alkaline-like ions (Ca$^+$, Sr$^+$, Yb$^+$).
This makes the transition frequency insensitive to the electric field gradients imposed by adjacent ions.
With direct readout available, In$^+$ is thus ideal for a multi-ion clock \cite{Tanja2019}, which can overcome the inherent stability limitations of a single-ion system \cite{Itano1993}.
The variety of optical clock systems has greatly stimulated progress in this field of research, and new proposals even consider methods to maintain a flexible and competitive approach beyond a redefinition of the SI second \cite{Lodewyck2019}.

Despite the appeal of the $^{115}$In$^+$ ion for a high-performance optical clock, its adoption has been hampered by concerns about technical reliability, particularly in the operation of deep UV laser systems.
Consequently, even the determination of the clock frequency \cite{vonZanthier00,WANG2007} has only been achieved recently.
Our own experiments find the In$^+$ clock to be robust and reliable.
After previous spectroscopic measurements \cite{OE2017}, we recently achieved clock operation by direct laser stabilization to the clock transition \cite{HI2019}.
Here we report the first frequency ratio measurement comparing the $^{115}$In$^+$ clock to another optical clock with an uncertainty of 7.7$\times$10$^{-16}$, which we consider a further step toward demonstrating the performance potential.

The In$^+$ optical clock probes the $^1$S$_0$--$^3$P$_0$ transition at 237~nm in an ion sympathetically cooled with a $^{40}$Ca$^+$ ion in a linear trap.
The clock laser pulses are generated by two-stage frequency doubling of a diode laser at 946~nm stabilized to an ultra-low expansion (ULE) cavity.
The fundamental frequency is steered such that the quadrupled frequency probes the $^1$S$_0$, $m_{\rm{F}} = \pm 9/2$ to $^3$P$_0$, $m_{\rm{F}} = \pm 7/2$ transitions. 
The fundamental laser is converted by second-harmonic generation first in a periodically-poled potassium titanyl phosphate (PPKTP) crystal and then a beta barium borate (BBO) crystal, installed in independent enhancement cavities.
The surface of the BBO crystal is AR coated for 473~nm and 237~nm.
Although a Brewster-cut crystal in a dry environment is usually employed to avoid degradation from UV-induced coating damage \cite{Hannig2018}, our BBO crystal operates in air without significant loss of efficiency.
After 6 years of operation, we find the laser systems to be no more problematic than those of other clocks.

The ratio measurement is performed using the $^{87}$Sr optical lattice clock NICT-Sr1 \cite{Hachisu2018}, located in the same building.
It probes approximately 1000 Sr atoms loaded into a vertically oriented one-dimensional optical lattice after two-stage laser cooling and presently achieves a systematic uncertainty of 7.3$\times$10$^{-17}$.

\begin{figure}[tbh]
\centering
\includegraphics[width=\linewidth]{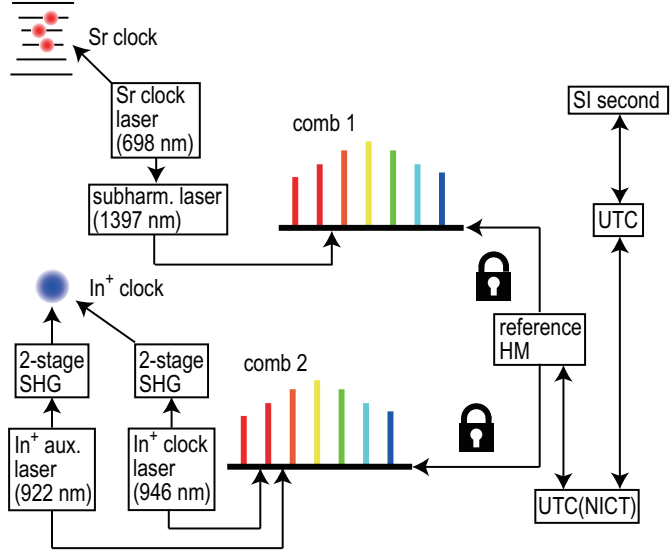}
\caption{Schematic representation of the frequency ratio measurement. 
The optical frequencies of the In$^+$ clock and the Sr clock are recorded as beat notes with independent optical frequency combs locked to a common hydrogen maser (HM). Comb 2 is also used to stabilize the frequency of an auxiliary laser. SHG; second harmonic generation.
}
\label{fig:optical_setup}
\end{figure}

The schematic representation of the frequency ratio measurement is depicted in Fig. \ref{fig:optical_setup}.
The measurement employs two independent Er-doped fiber frequency combs. One of them is configured to perform long-term measurements of the Sr clock transition frequency, using a subharmonic laser at 1397 nm to generate a comb beat signal (comb 1).
The other (comb 2) is equipped with spectral broadening and frequency doubling systems for generating optical beat signals with the fundamental wavelengths of 946~nm 
used by the $^{115}$In$^+$ clock laser and 922~nm for an auxiliary laser employed in state preparation and detection, where the comb provides long term stability.

With repetition rate $f_{\rm{r}}$ and carrier-envelope offset $f_{\rm{CEO}}$ phase-locked to signals derived from a hydrogen maser (HM), the comb line frequencies $f_n=n \times f_{\rm{r}}+f_{\rm{CEO}}$ of both combs faithfully reproduce the behavior of the HM microwave signal within the optical regime.
It is convenient to describe frequencies $f=f_0(1+y)$ in terms of a small fractional deviation $y$ from a nominal value $f_0$. 
When all frequency offsets and multiplications are accounted for in the calculation of $f_0$, the small residual $y$ is independent of frequency.
A HM deviation $y_{\rm{HM}}$ is then equally reproduced in the comb lines serving as reference to the two clocks,
where they appear as $\Delta y_{\rm{In}}=y_{\rm{In}}-y_{\rm{HM}}$ and $\Delta y_{\rm{Sr}}=y_{\rm{Sr}}-y_{\rm{HM}}$.
In the absence of excess noise, the HM contribution is eliminated in the difference 
$\Delta y_{\rm{In}}-\Delta y_{\rm{Sr}}=y_{\rm{In}}- y_{\rm{Sr}}\simeq R/R_0-1$.
Here $R= f_{\rm{In}}/f_{\rm{Sr}}$ is the desired frequency ratio, while $R_0= f_{\rm{In_0}}/f_{\rm{Sr_0}}$ is the ratio of nominal values where we adopt the CIPM recommended frequencies of 2017.
The HM deviation is typically of order 10$^{-12}$, such that the approximation yields no loss in accuracy.

\begin{figure}[tbh]
\centering
\includegraphics[width=\linewidth]{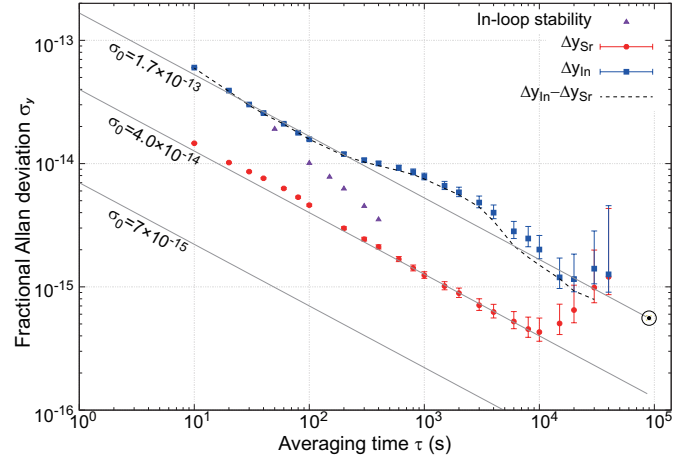}
\caption{Fractional Allan deviations of $\Delta y_{\rm{In}}$ (blue squares) and $\Delta y_{\rm{Sr}}$ (red circles).
We find the In$^{+}$ statistical uncertainty $u_{\rm{a}}^{\rm{In}}=5.6 \times10^{-16}$ by extrapolating to the full measurement time of 89 445 s (circular mark).
All data has been concatenated into a single interval, and the rising instability for $\tau >$ 10 000 s represents the HM drift over four successive days of operation.
We calculate the stability of the frequency ratio as represented by $\Delta y_{\rm{In}}-\Delta y_{\rm{Sr}}$ for 75 381 s of simultaneous clock operation (dashed line).
While the cancellation of the HM contribution only results in a minor improvement in short-term instability, the ratio is not affected by the long-term drift. The purple triangles represent the in-loop stability of the In$^+$ clock servo, starting at $2 \times 10^{-14}$ at the 50 s steering interval.
Solid guide lines represent instabilities $\sigma_y(\tau) = \sigma_0 / \sqrt{\tau / \rm{s}}$ as labeled, where $\sigma_0=7\times10^{-15}$ represents the stability of the Sr clock.
}
\label{fig:fig3}
\end{figure}

To characterize the stability, we calculate the overlapping Allan deviation for $\Delta y_{\rm{In}}$, $\Delta y_{\rm{Sr}}$ and $\Delta y_{\rm{In}} - \Delta y_{\rm{Sr}}$ as shown in Fig. \ref{fig:fig3}.
The dominant source of instability is the In$^+$ clock and its measurement.
We observe an overall white-frequency-noise limited behavior characterized by $\sigma_ y(\tau) = 1.7 \times 10^{-13}/ \sqrt{\tau / \rm{s}}$, overlaid by excess instability around $\tau = 10^3$ s.
We attribute this to the measurement by comb 2, which implements a frequency counting system more susceptible to phase instabilities: Whereas comb 1 detects a 8.2~GHz harmonic of the repetition rate referenced to a 100~MHz HM signal, comb 2 detects a 1~GHz harmonic referenced to 10~MHz.
The Allan deviation supports such effects averaging out over longer measurements, with negligible effect on the mean result.

\begin{figure}[htbp]
\centering
\includegraphics[width=\linewidth]{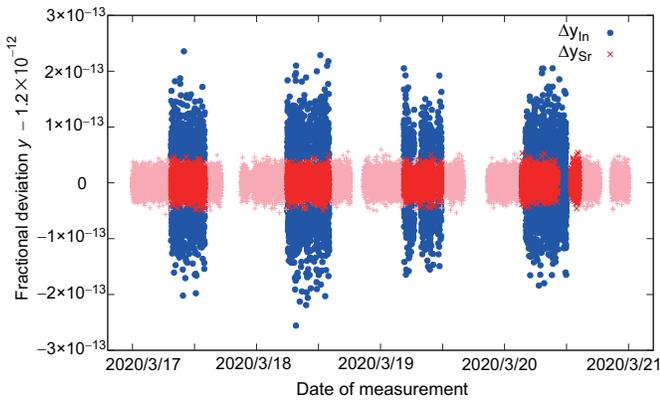}
\caption{Frequency measurements of In$^+$ and Sr clocks. Results are shown as fractional deviation after subtracting a common baseline of 1.2$\times 10^{-12}$ to accommodate the mean HM frequency $\overline{y_{\rm{HM}}}$=1.2001$\times 10^{-12}$. The blue '$\bullet$' symbols represent $\Delta y_{\rm{In}}$ while the red '$\times$' symbols represent the evaluated subset of $\Delta y_{\rm{Sr}}$. The pink '$+$' symbols represent $\Delta y_{\rm{Sr}}$ data which does not contribute to the ratio calculation. All symbols represent averages over 10~s intervals, incomplete intervals are omitted from the plot.
}
\label{fig:fig2}
\end{figure}

We calculate a mean value $\overline{\Delta y_{\rm{In}}}$ over all available In$^+$ data, with a statistical uncertainty of $u_{\rm{a}}^{\rm{In}} = 5.6 \times 10^{-16}$ determined by extrapolating $\sigma_y(\tau)$ to the full duration $\tau=89~445$ s.
$\overline{\Delta y_{\rm{Sr}}}$ is evaluated in the same way, using a similarly long subset of the available data (Fig. \ref{fig:fig2}) that is selected according to three criteria:
\begin{description}
\item [(a)]	All simultaneously acquired data is included. For this data, any HM instability affects $\Delta y_{\rm{In}}$ and $\Delta y_{\rm{Sr}}$ identically and cancels in the determined ratio.
\item [(b)]	The temporal barycenter of the Sr subset differs from that of the In$^+$ distribution by less than the 10 s length of the evaluation blocks. Over this time the drift of the HM is fully negligible.
\item [(c)]	The distribution of the In$^+$ data is reproduced as closely as possible. The HM serves as a flywheel oscillator in the comparison of the two intervals. Minimizing the time to extrapolate over minimizes the uncertainty.
\end{description}
We find $u_{\rm{a}}^{\rm{Sr}} = 1.3 \times 10^{-16}$ as the statistical uncertainty over 89~460~s of Sr data.
The effect of stochastic HM behavior during the non-overlapping intervals of the two distributions is handled by the same model applied in frequency comparisons of the Sr clock to the international time scale \cite{Nemitz2020}. For the selected Sr data, this contributes $u_{\rm{a}}^{\rm{HM}} = 8.5 \times 10^{-17}$ to the overall statistical uncertainty of $u_{\rm{a}} ^2= (u_{\rm{a}}^{\rm{In}}) ^2+ (u_{\rm{a}}^{\rm{Sr}})^2 + (u_{\rm{a}}^{\rm{HM}})^2= (5.8\times 10^{-16})^2$.
We consider this to be a slightly conservative value neglecting the cancellation of short-term HM noise. However, even a perfect cancellation of noise according to $u_{\rm{a}}^{\rm{Sr}}=1.3 \times 10^{-16}$ would not reduce the statistical uncertainty to less than $5.5 \times 10^{-16}$.
The Allan deviation of $\Delta y_{\rm{In}}-\Delta y_{\rm{Sr}}$  for 75 381 s of simultaneous data confirms that drift and similarly slow variations of the HM frequency are cancelled in the determination of the frequency ratio. 

The systematic uncertainty evaluation of the $^{115}$In$^+$ clock is largely identical to our previous report \cite{OE2017}.
However, where the clock frequency was previously determined from spectra obtained by laser scans across transitions that did not resolve near-degenerate Zeeman sublevels,
it is now determined by actively stabilizing the clock laser to represent the center of two resolved transitions between Zeeman sublevels ($^1$S$_0$, $m_{\rm{F}}=\pm9/2$ to $^3$P$_0$, $m_{\rm{F}}=\pm7/2$).
A dynamic application of magnetic field throughout the clock laser irradiation periods \cite{HI2019} lifts the degeneracy between the Zeeman sublevels.
This simultaneously allows precise determination of the magnetic field using the known $g$-factors of the upper and lower energy levels.
The quadratic Zeeman shift is estimated to be 8 mHz using the magnetic field value and the known coefficient \cite{Tanja2019}.
Another improvement is a reduction of the 2nd order Doppler shift due to secular motion.
The auxiliary laser derived from a fundamental wavelength of 922~nm addresses the $^1$S$_0$--$^3$P$_1$ transition at 230~nm.
Normally used for optical pumping and state readout, we inserted a pulse of this laser for direct cooling of the In$^+$--Ca$^+$ motion along the linear trap axis just before the clock interrogation period.
The natural linewidth of 360~kHz allows a much lower temperature than the cooling of the $^{40}$Ca$^+$ ion on the 22 MHz wide $^2$S$_{1/2}$--$^2$P$_{1/2}$ transition.
The temperature of the in-phase mode estimated from the ratio of the blue and red sidebands \cite{Tanaka2015} is then reduced to about 70 $\mu \rm{K}$.
Although the implementation of the cooling is currently limited to one degree of motional freedom, the uncertainty due to secular motion is reduced from 0.7~Hz to 0.5~Hz.
By installing the trap electrode in a different vacuum chamber and measuring its temperature under operating conditions, the BBR shift was estimated to be $-$34(1) mHz.
The systematic uncertainty budget is listed in table \ref{systematics}, with a total fractional uncertainty of $u_{\rm{b}}^{\rm{In}}=5.0\times10^{-16}$ corresponding to 0.6~Hz.
This is a nearly 8-fold reduction in systematic uncertainty compared to our previous measurement \cite{OE2017}, which was limited by a Zeeman shift uncertainty of 4.6 Hz.

The Sr clock contributes a systematic uncertainty of $u_{\rm{b}}^{\rm{Sr}} =7.3\times$10$^{-17}$.
Combined with $u_{\rm{a}}=5.8\times10^{-16}$,
this yields an overall fractional uncertainty of 7.7$\times 10^{-16}$, and we determine the optical frequency ratio to be $R=R_0(1+\overline{\Delta y_{\rm{In}}}-\overline{\Delta y_{\rm{Sr}}})=2.952\,748\,749\,874\,863\,3\,(23)$.

\begin{table}
\caption{Fractional shifts and uncertainties in $^{115}\rm{In}^+$ system as well as frequency link to the SI second.  BBR; blackbody radiation, S.M.; secular motion, M.M.; micromotion.}
\begin{center}
\begin{tabular}{lcc}
\hline
Effects & Shift  & Uncertainty \\
& $(10^{-16})$ & $(10^{-16})$ \\
\hline
2nd order Zeeman shift &  0.06 & $<0.1$ \\
Clock laser Stark shift & $-$0.03 & $<0.1$ \\
BBR shift & $-$0.27 &  $<0.1$\\
2nd order Doppler (S.M.) & $-$0.004 &  3.9\\
2nd order Doppler (M.M.) & 0 & 3.1\\
\hline
\bf{In$^+$ systematics} &\bf{ $-$0.24} &  \bf{5.0}       \\
In$^+$ statistics ($u_{\rm{a}}^{\rm{In}}$)& & 5.6 \\
\hline
\bf{In$^+$ total} & &  \bf{7.5}       \\
Gravitational shift & 83.11& 0.2 \\
HM extrapolation & & 2.4 \\
UTC(NICT)$-$UTC & & 2.3 \\
UTC extrapolation (25 days) & & 1.8 \\
UTC$-$SI second & & 1.3 \\
\hline
\bf{Overall total} & & \bf{8.5}
\end{tabular}
\end{center}
\label{systematics}
\end{table}

Since the reference HM is part of the Japan Standard Time System (JST) at NICT \cite{Hanado2007}, its frequency can also be traced to the SI second, allowing for an absolute measurement of the $^{115}$In$^+$ clock frequency.
This begins with the fractional frequency deviation of the international timescale UTC, determined by the International Bureau of Weights and Measures (BIPM), and published for the 35-day period of Feb 25 to Mar 31 2020 in volume 387 of its monthly Circular T. 
Based on measurements of multiple primary and secondary frequency standards, the reported value corresponds to a deviation $y_{\rm{UTC}}=5.2(1.3)\times 10^{-16}$ from the nominal frequency.
We then use the stability of the international timescale to extrapolate this value to a shorter 25-day interval, and determine the frequency difference of UTC(NICT) relative UTC from the time delays published in the same Circular T.
From the 25-day interval, the stability of the JST HM ensemble allows us to find the mean frequency of the reference HM for the time of the In$^+$ measurements with $2.4\times10^{-16}$ additional uncertainty \cite{Nemitz2020}.
In this remote comparison, the height of the In$^+$ clock relative to the geoid introduces a gravitational redshift of 83.11(2)$\times 10^{-16}$.
Combining these uncertainties yields $u_{\rm{link}} = 4.0\times 10^{-16}$  for the uncertainty in the link to the SI second through the international timescale.
Table \ref{systematics} lists the individual uncertainty contributions along with the overall uncertainty.
The absolute $^{115}$In$^+$ clock frequency is determined as 1~267~402~452~901~040.1(1.1)~Hz.
With a fractional uncertainty of $8.5\times10^{-16}$, this six-fold reduction represents significant improvement in both systematic and statistical uncertainties over our last report.
In relation to the previous value of 1~267~402~452~901~049.9~(6.9)~Hz \cite{OE2017},
the new result differs by about $1.4 \sigma$.
While the source of the discrepancy is not clearly identified, the possibility of errors from spectral distortions has been the driving motivation in implementing the new scheme of field application during spectroscopy, which eliminates this source of error.

To confirm the link to the SI second, we can calculate the $^{115}$In$^+$ frequency from $R$ and the CIPM recommendation for $f_{\rm{Sr}}$ as $f'_{\rm{In}}=R \times f_{\rm{Sr}}$ to find 1~267~402~452~901~040.0(1.1)~Hz, with a relative uncertainty of 8.6$\times$10$^{-16}$.
The difference of the $^{115}$In$^+$ clock frequencies determined in these ways is 0.087 Hz ($6.9\times10^{-17}$), and it is well within the additional uncertainty of the frequency link to the SI second.

While the Doppler shifts from ion motion limit this first frequency ratio measurement involving the $^{115}$In$^+$ clock to a fractional uncertainty of $7.7\times10^{-16}$, extending the motional cooling scheme to all degrees of freedom 
together with a more accurate evaluation of the residual micromotion will further reduce the uncertainty.
Other uncertainty contributions represent the statistical uncertainty of in-situ measurements.
These, along with the statistical contribution in any clock comparison, are expected to dramatically improve with increased stability of the clock laser.
An additional improvement in clock stability would result from increased duty cycle.
Where the current measurements only probe the clock transition for less than 20 \% of the cycle time, an optical system that makes more efficient use of the collected photons will decrease the time dedicated to state detection.
A new optical frequency comb system is already in preparation to reduce measurement instability and allow direct optical comparisons.
With future improvements, we hope to demonstrate the $^{115}$In$^+$ clock as a competitor to established candidates \cite{Huntemann2016,Brewer2019}, combining the insensitivity to environmental effects found in $^{27}$\!Al$^+$ with an interrogation scheme that does not require the complexities of quantum logic techniques.


\bibliography{Ohtsubo_ref}

\begin{thebibliography}{10}
\newcommand{\enquote}[1]{``#1''}

\bibitem{Hong_2016}
F.-L. Hong, \enquote{Optical frequency standards for time and length
  applications,} {\protect\JournalTitle{Measurement Science and Technology}}
  \textbf{28}, 012002 (2016).

\bibitem{Riehle2018}
F.~Riehle, P.~Gill, F.~Arias, and L.~Robertsson, \enquote{The {CIPM} list of
  recommended frequency standard values: guidelines and procedures,}
  {\protect\JournalTitle{Metrologia}} \textbf{55}, 188--200 (2018).

\bibitem{Lodewyck2019}
J.~Lodewyck, \enquote{On a definition of the {SI} second with a set of optical
  clock transitions,} {\protect\JournalTitle{Metrologia}} \textbf{56}, 055009
  (2019).

\bibitem{Rosenband2008}
T.~Rosenband, D.~B. Hume, P.~O. Schmidt, C.~W. Chou, A.~Brusch, L.~Lorini,
  W.~H. Oskay, R.~E. Drullinger, T.~M. Fortier, J.~E. Stalnaker, S.~A. Diddams,
  W.~C. Swann, N.~R. Newbury, W.~M. Itano, D.~J. Wineland, and J.~C. Bergquist,
  \enquote{Frequency ratio of {A}l$^+$ and {H}g$^+$ single-ion optical clocks;
  metrology at the 17th decimal place,} {\protect\JournalTitle{Science}}
  \textbf{319}, 1808--1812 (2008).

\bibitem{Godun2014}
R.~M. Godun, P.~B.~R. Nisbet-Jones, J.~M. Jones, S.~A. King, L.~A. M.~H.
  Johnson, S.~Margolis, K.~Szymaniec, S.~N. Lea, K.~Bongs, and P.~Gill,
  \enquote{Frequency ratio of two optical clock transitions in $^{171}${Y}b$^+$
  and constraints on the time variation of fundamental constants,}
  {\protect\JournalTitle{Phys. Rev. Lett.}} \textbf{113}, 210801 (2014).

\bibitem{Matsubara2012}
K.~Matsubara, H.~Hachisu, Y.~Li, S.~Nagano, C.~Locke, A.~Nogami, M.~Kajita,
  K.~Hayasaka, T.~Ido, and M.~Hosokawa, \enquote{Direct comparison of a
  {C}a$^+$ single-ion clock against a {S}r lattice clock to verify the absolute
  frequency measurement,} {\protect\JournalTitle{Optics Express}} \textbf{20},
  22034--22041 (2012).

\bibitem{Tyumenev2016}
R.~Tyumenev, M.~Favier, S.~Bilicki, E.~Bookjans, R.~Le~Targat, J.~Lodewyck,
  D.~Nicolodi, Y.~Le~Coq, M.~Abgrall, J.~Guena, L.~De~Sarlo, and S.~Bize,
  \enquote{Comparing a mercury optical lattice clock with microwave and optical
  frequency standards,} {\protect\JournalTitle{New J. Phys.}} \textbf{18},
  113002 (2016).

\bibitem{Nemitz2016}
N.~Nemitz, T.~Ohkubo, M.~Takamoto, I.~Ushijima, M.~Das, N.~Ohmae, and
  H.~Katori, \enquote{Frequency ratio of {Y}b and {S}r clocks with 5x10$^{-17}$
  uncertainty at 150 s averaging time,} {\protect\JournalTitle{Nat. Photon.}}
  \textbf{10}, 258--264 (2016).

\bibitem{Ohmae2020}
N.~Ohmae, F.~Bregolin, N.~Nemitz, and H.~Katori, \enquote{Direct measurement of
  the frequency ratio for {H}g and {Y}b optical lattice clocks and closure of
  the {H}g/{Y}b/{S}r loop,} {\protect\JournalTitle{Optics Express}}
  \textbf{28}, 15112--15121 (2020).

\bibitem{BACON2020}
K.~Beloy, M.~I. Bodine, T.~Bothwell, S.~M. Brewer, S.~L. Bromley, J.-S. Chen,
  J.-D. Desch$\rm{\hat{e}}$nes, S.~A. Diddams, R.~J. Fasano, T.~M. Fortier,
  Y.~S. Hassan, D.~B. Hume, D.~Kedar, C.~J. Kennedy, I.~Khader, A.~Koepke,
  D.~R. Leibrandt, H.~Leopardi, A.~D. Ludlow, W.~F. McGrew, W.~R. Milner, N.~R.
  Newbury, D.~Nicolodi, E.~Oelker, T.~E. Parker, J.~M. Robinson, S.~Romisch,
  S.~A. Schäffer, J.~A. Sherman, L.~C. Sinclair, L.~Sonderhouse, W.~C. Swann,
  J.~Yao, J.~Ye, and X.~Zhang, \enquote{Frequency ratio measurements with
  18-digit accuracy using a network of optical clocks,}  (2020).
  ArXiv:2005.14694.

\bibitem{Dehmelt1983}
H.~G. Dehmelt, \enquote{Monoion oscillator as potential ultimate laser
  frequency standard,} {\protect\JournalTitle{IEEE Transactions on
  Instrumentation and Measurement}} pp. 83--87 (1982).

\bibitem{Brewer2019}
S.~M. Brewer, J.-S. Chen, A.~M. Hankin, E.~R. Clements, C.~W. Chou, D.~J.
  Wineland, D.~B. Hume, and D.~R. Leibrandt, \enquote{$^{27}${A}l$^{+}$
  quantum-logic clock with a systematic uncertainty below
  ${10}^{\ensuremath{-}18}$,} {\protect\JournalTitle{Phys. Rev. Lett.}}
  \textbf{123}, 033201 (2019).

\bibitem{Tanja2019}
J.~Keller, T.~Burgermeister, D.~Kalincev, A.~Didier, A.~P. Kulosa, T.~Nordmann,
  J.~Kiethe, and T.~E. Mehlst\"{a}ubler, \enquote{Controlling systematic
  frequency uncertainties at the 10$^{-19}$ level in linear coulomb crystals,}
  {\protect\JournalTitle{Phys. Rev. A}} \textbf{99}, 013405 (2019).

\bibitem{Itano1993}
W.~M. Itano, J.~C. Bergquist, J.~J. Bollinger, J.~M. Gilligan, D.~J. Heinzen,
  F.~L. Moore, M.~G. Raizen, and D.~J. Wineland, \enquote{Quantum projection
  noise: Population fluctuations in two-level systems,}
  {\protect\JournalTitle{Phys. Rev. A}} \textbf{47}, 3554--3570 (1993).

\bibitem{vonZanthier00}
J.~von Zanthier, T.~Becker, M.~Eichenseer, A.~Y. Nevsky, C.~Schwedes, E.~Peik,
  H.~Walther, R.~Holzwarth, J.~Reichert, T.~Udem, T.~W. H\"{a}nsch, P.~V.
  Pokasov, M.~N. Skvortsov, and S.~N. Bagayev, \enquote{Absolute frequency
  measurement of the {I}n$^+$ clock transition with a mode-locked laser,}
  {\protect\JournalTitle{Opt. Lett.}} \textbf{25}, 1729--1731 (2000).

\bibitem{WANG2007}
Y.~Wang, R.~Dumke, T.~Liu, A.~Stejskal, Y.~Zhao, J.~Zhang, Z.~Lu, L.~Wang,
  T.~Becker, and H.~Walther, \enquote{Absolute frequency measurement and high
  resolution spectroscopy of $^{115}$in$^+$ 5s$^2$ $^1${S}$_0$–5s5p
  $^3${P}$_0$ narrowline transition,} {\protect\JournalTitle{Optics
  Communications}} \textbf{273}, 526 -- 531 (2007).

\bibitem{OE2017}
N.~Ohtsubo, Y.~Li, K.~Matsubara, T.~Ido, and K.~Hayasaka, \enquote{Frequency
  measurement of the clock transition of an indium ion sympathetically-cooled
  in a linear trap,} {\protect\JournalTitle{Optics Express}} \textbf{25},
  11725--11735 (2017).

\bibitem{HI2019}
N.~Ohtsubo, Y.~Li, N.~Nemitz, H.~Hachisu, K.~Matsubara, T.~Ido, and
  K.~Hayasaka, \enquote{Optical clock based on a sympathetically-cooled indium
  ion,} {\protect\JournalTitle{Hyperfine Interact.}} \textbf{240}, 39 (2019).

\bibitem{Hannig2018}
S.~Hannig, J.~Mielke, J.~A. Fenske, M.~Misera, N.~Beev, C.~Ospelkaus, and P.~O.
  Schmidt, \enquote{A highly stable monolithic enhancement cavity for second
  harmonic generation in the ultraviolet,} {\protect\JournalTitle{Review of
  Scientific Instruments}} \textbf{89}, 013106 (2018).

\bibitem{Hachisu2018}
H.~Hachisu, F.~Nakagawa, Y.~Hanado, and T.~Ido, \enquote{Months-long real-time
  generation of a time scale based on an optical clock,}
  {\protect\JournalTitle{Sci. Rep.}} \textbf{8}, 4243 (2018).

\bibitem{Nemitz2020}
N.~Nemitz, T.~Gotoh, F.~Nakagawa, H.~Ito, Y.~Hanado, T.~Ido, and H.~Hachisu,
  \enquote{Absolute frequency of $^{87}${S}r at $1.8\times10^{-16}$ uncertainty
  by reference to remote primary frequency standards,}  (2020).
  ArXiv:2008.00723.

\bibitem{Tanaka2015}
U.~Tanaka, T.~Kitanaka, K.~Hayasaka, and S.~Urabe, \enquote{Sideband cooling of
  a {C}a$^+$–{I}n$^+$ ion chain toward the quantum logic spectroscopy of
  {I}n$^+$,} {\protect\JournalTitle{Applied Physics B}} \textbf{121}, 147–153
  (2015).

\bibitem{Hanado2007}
Y.~Hanado, K.~Imamura, N.~Kotake, F.~Nakagawa, Y.~Shimizu, R.~Tabuchi,
  Y.~Takahashi, M.~Hosokawa, and T.~Morikawa, \enquote{{T}he {N}ew {G}eneration
  {S}ystem of {J}apan {S}tandard {T}ime at {NICT},}
  {\protect\JournalTitle{International Journal of Navigation and Observation}}
  \textbf{2008}, 841672 (2007).

\bibitem{Huntemann2016}
N.~Huntemann, C.~Sanner, B.~Lipphardt, C.~Tamm, and E.~Peik,
  \enquote{Single-ion atomic clock with
  $3\ifmmode\times\else\texttimes\fi{}{10}^{\ensuremath{-}18}$ systematic
  uncertainty,} {\protect\JournalTitle{Phys. Rev. Lett.}} \textbf{116}, 063001
  (2016).

\end{thebibliography}






\end{document}